\documentclass[superscriptaddress,twocolumn,showpacs,prb]{revtex4-2}
\usepackage[utf8]{inputenc} 
\usepackage{amsmath}
\usepackage{braket}
\usepackage{graphicx}
\usepackage{amsfonts}
\usepackage{pgfplots}
\pgfplotsset{compat=newest}
\usepackage{csquotes}
\usepackage{hhline}
\usepackage{amssymb}
\usepackage{listings}
\usepackage{color}
\usepackage{amssymb}
\usepackage{amsfonts} 
\usepackage{float}

\DeclareMathAlphabet{\mathcal}{OMS}{cmsy}{m}{n}
\SetMathAlphabet{\mathcal}{bold}{OMS}{cmsy}{b}{n}

\definecolor{codegreen}{rgb}{0,0.6,0}
\definecolor{codegray}{rgb}{0.5,0.5,0.5}
\definecolor{codepurple}{rgb}{0.58,0,0.82}
\definecolor{backcolour}{rgb}{0.95,0.95,0.92}
\lstdefinestyle{mystyle}{
    backgroundcolor=\color{backcolour},   
    commentstyle=\color{codegreen},
    keywordstyle=\color{magenta},
    numberstyle=\tiny\color{codegray},
    stringstyle=\color{codepurple},
    basicstyle=\footnotesize,
    breakatwhitespace=false,         
    breaklines=true,                 
    captionpos=b,                    
    keepspaces=true,                 
    numbers=left,                    
    numbersep=5pt,                  
    showspaces=false,                
    showstringspaces=false,
    showtabs=false,                  
    tabsize=2
}
 
\usepackage{dcolumn}
\usepackage{tabularx}
\usepackage[colorlinks=true,linkcolor=blue,citecolor=blue,urlcolor=blue,draft=false]{hyperref}
\usepackage{longtable}
\usepackage{braket}
\usepackage{float}
\newcolumntype{C}{>{\centering\arraybackslash}X}
\usepackage{afterpage}
\usepackage{textcomp}

\begin{document}
\title{Simulation of Lennard-Jones Potential on a Quantum Computer}

\author{Prabhat}
\email{prabhat.70707@gmail.com}
\affiliation{Physics Department, \\St.Stephen's college, University of Delhi, Sudhir Bose Marg, University Enclave, New Delhi, Delhi 110007, India}
\author{Bikash K. Behera}
\email{bikash@bikashsquantum.com}
\affiliation{Bikash's Qauntum (OPC) Pvt. Ltd., Balindi, Mohanpur 741246, Nadia, West Bengal, India}
\vspace{5 cm}

\begin{abstract}
Simulation of time dynamical physical problems has been a challenge for classical computers due to their time-complexity. To demonstrate the dominance of quantum computers over classical computers in this regime, here we simulate a semi-empirical model where two neutral particles interact through Lennard-Jones potential in a one dimensional system. We implement the above scenario on IBM quantum experience platform using a 5-qubit real device. We construct the Hamiltonian and then efficiently map it to quantum operators onto quantum gates using the time-evolutionary unitary matrix obtained from the Hamiltonian. We verify the results collected from the qasm-simulator and compare it with that of the 5-qubit real chip ibmqx2.
\end{abstract}

\begin{keywords}{Quantum Simulation, Time-Complexity, IBM Quantum Experience, Quantum Gates, Schr$\o$dinger's Equation}\end{keywords}

\maketitle
\section{Introduction}
The age of quantum simulation on quantum computers started in the 1980s when Benioff showed that a computer could operate under the laws of quantum mechanics, which is further backed with the famous paper by Feynman at the first conference on the physics of computation, held at MIT in May 1980. He presented in the talk, titled as `Quantum mechanical Hamiltonian models of discrete processes that erase their histories: application to Turing machines', how efficient can a quantum computer be for NP Hard and EXP problems \cite{t-comp} and there proposed a basic model for a quantum computer \cite{feynmen:1982}. The problem of simulating the full-time evolution of arbitrary quantum systems is intractable due to the exponential growth of the size of a system. For example, it would need $2^{47}$ bits to record a state of 47 spin-\textonehalf particles, which is equal to $1.4 \times 10^{14}$ bits, a little less than 20 TB which is the largest storage capacity of hard disk currently \cite{ hd0, hd1, hd2}. Feynman gave simple examples of one quantum system simulating another and conjectured that there existed a class of universal quantum simulators capable of simulating any quantum system that evolved according to local interactions \cite{lloyd}. These local interactions, if controlled, could be exploited to mimic the dynamics of the quantum systems very efficiently. Essentially, each tweaking will move the system in specific directions in Hilbert space which corresponds to an operation, we are mimicking through it.

Classical information was first carefully defined by Shannon in his paper in 1948 \cite{qnm_ShannonIEEE1948}. Any change could be considered as a signal if it can be encoded with some data by the sender and retrieved by the receiver. `Bit' is the most popular unit of classical data. A bit can either store 1 or 0 in form of voltages generally. However, in simple terms, due to the quantum nature of the qubit (quantum bit), it can store more information than a bit. A qubit is an abstraction of any two-state quantum system, like a bit, is an abstraction of a two-state classical system. Information can be encoded on photons, spins, atoms, or the energy state of electrons in quantum dots. A gate is the function of bits. Since a qubit is more versatile than a bit (can be in a state, which is a combination of its two states simultaneously, is one of them), there are many more fundamental gates. These gates control the output and a series of them can perform any complicated tasks. Though now the number of qubits required will be 47 (exponentially small), the number of operations is still the same to evolve the system through time. It is a branch of computer science, physics, and engineering in its infant stages, growing rapidly. The quest for more QS is described in referred works \cite{ quantumsim5, quantumsim6, quantumsim7, quantumsim8, quantumsim9, quantumsim10, quantumsim11, quantumsim13, quantumsim14, quantumsim15, quantumsim17, quantumsim18, quantumsim19, quantumsim20, quantumsim21}. Currently, using an online accessible quantum computing platform, IBM quantum experience, a number of research works have been performed among which the following \cite{ ibm, ibm2, ibm3, ibm4, ibm5, ibm6, ibm7, ibm8, ibm9, ibm10, ibm11, ibm12, ibm13, ibm14, ibm15, ibm16, ibm17, ibm18, ibm19, ibm20, ibm21, ibm22, ibm23, ibm24, ibm25}
can be referred.

The organization of the paper is as follows. In Section \ref{qsp_sec2}, the theoretical background and the algorithm is laid out. Then the mapping is given in Sec. \ref{qsp_sec3}. The simulation is then done using the Qiskit of IBM quantum experience platform. Finally, the experimental results of simulation and comparisons are given in Sec. \ref{qsp_sec4}. The analysis and future scope is discussed in the Sec. \ref{qsp_sec5}.

\section{Simulation \label{qsp_sec2}}
\begin{figure}[h]
\centering
    \includegraphics[scale=0.4]{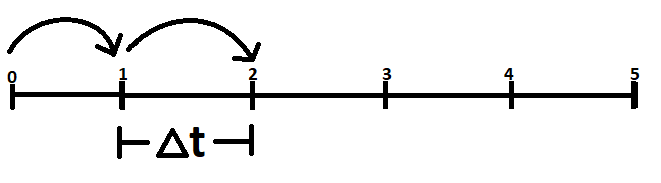}
    \caption{The system will evolve through time into discrete time stages. Apart from the discontinuity in time, we have discrete spatial points too. As long as two black boxes give out the same output to the same input, we regard them as identical. This is the philosophical idea behind simulations. Though the mechanism could be different in both the cases, real-world and computer, we can consider them to be same machinery.}
\label{time-step}
\end{figure}

Since on a computer, the notion of infinitesimal change can not be applied. Thus, as a numerical approximation, the system's evolution has been studied discretely (Fig. \ref{time-step}). We can always make our time step sufficiently small, for depicting it as close as to a physical system. But there is always a limit to this. This limit depends on the ability of our computer to handle these numbers, the runtime one can afford and the precision one wants in the system. Therefore, only the discretized system and its operators are being considered in subsequent sections.

\begin{figure}[h]
\centering
\includegraphics[scale=0.4]{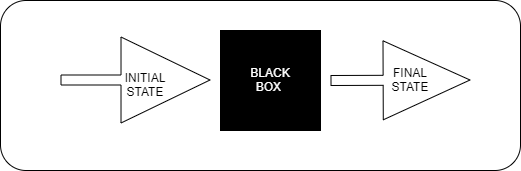}
\caption{We can construct perfect laws if we have all the inputs and their corresponding outputs. Given the information in this black box, you can predict the future state of your system depending upon the current state.}
\label{digital simulation}
\end{figure}

It is said that, given the dynamical law (Eq. \ref{digital simulation}) of a system, the whole state-space can be determined for given initial conditions \cite{theoriticalminimum}. To mimic this, the digital quantum simulations work on the following main steps: initial state preparation, time evolution, and measurement. Consequently, we designed our simulation based on that principle.

The main algorithm is focused on designing this time evolution of the system since that defines the system, and the other two parts are computational. Quantum systems evolve according to a unitary operator which is given as $U(t)=e^{-\iota \hat{H} t/\hbar}$ \cite{unitary}. The $\hat{H}$ is called the Hamiltonian of the system and is dictated by the components of the system. This unitary operation can be split into free and interacting operators using Trotter's formula \cite{trot}.

\begin{align*}
    exp(-\iota \Hat{H} t) &= exp( -\iota (\Hat{V}+\Hat{K}) t ) \\
                          &= exp( -\iota \Hat{V} t )exp( -\iota \Hat{K} t )exp( \mathcal{O} (\Delta t^2))
\end{align*}

We have used first order truncation but higher order approximation methods have also been developed \cite{ high0, high1, high2} that gives more accurate results at the cost of more number of quantum gates per time step.

\subsection{Hamiltonian}

The time-dependent Schrodinger wave equation in one dimension is

\begin{equation}
    \iota \hbar \frac{\partial}{\partial t}\Psi = \hat{H} \Psi
\end{equation}

where $\Psi$ is the time dependent wave-function that contain all the information about a system and $\hat{H}$ is the Hamiltonian of the system. The general form of Hamiltonian contains the operations associated with the kinetic and potential energies. Hence, for a wavefunction of two particles of mass $m_1$ and $m_2$ in one dimension can be written as,

\begin{equation}
    \hat{H}=\frac{-\hbar^2}{2m_1}\frac{\partial^2}{\partial x_1^2}+\frac{-\hbar^2}{2m_2}\frac{\partial^2}{\partial x_2^2} + V(x)
\end{equation}

Here, $\iota \hbar\frac{\partial}{\partial x}$ is the momentum operator for a wavefunction. The particles will be interacting through a short range force defined by Lennard-Jones potential \cite{bljp0,bljp1,bljp2,bljp3,bljp4,bljp5,bljp6,bljp7} which is given by 
\begin{equation}
    V(x) = \epsilon[\big(\frac{\sigma}{x} \big) ^{12} - \big( \frac{\sigma}{x} \big) ^6]
\end{equation}

where $\epsilon$ and $\sigma$ are constants for the particles in consideration and $x$ is the relative distance between them. To get rid of the unnecessary clutter, we took $\epsilon = 1/4$ and $\sigma = 1$ and $m_1=m_2 = 1/2$. This potential has two parts. The negative part has a larger range and responsible for the attraction, and the positive part has a shorter range and is responsible for the repulsion. It is widely used as a model for intermolecular interaction which results in a type of ideal fluids called Lennard Jones fluid \cite{ljf}. Therefore with this information, our Hamiltonian becomes (in abstract form)

\begin{equation}
    \widehat{H} = \hat{p}_{1}^{2} + \hat{p}_{2}^{2} + (\hat{x}^{-12} -  \hat{x}^{-6})
\end{equation}

where $\hat{p}$ and $\hat{x}$ are the momentum and position operators respectively.

\begin{figure}[h]
\centering
    \includegraphics[scale=0.4]{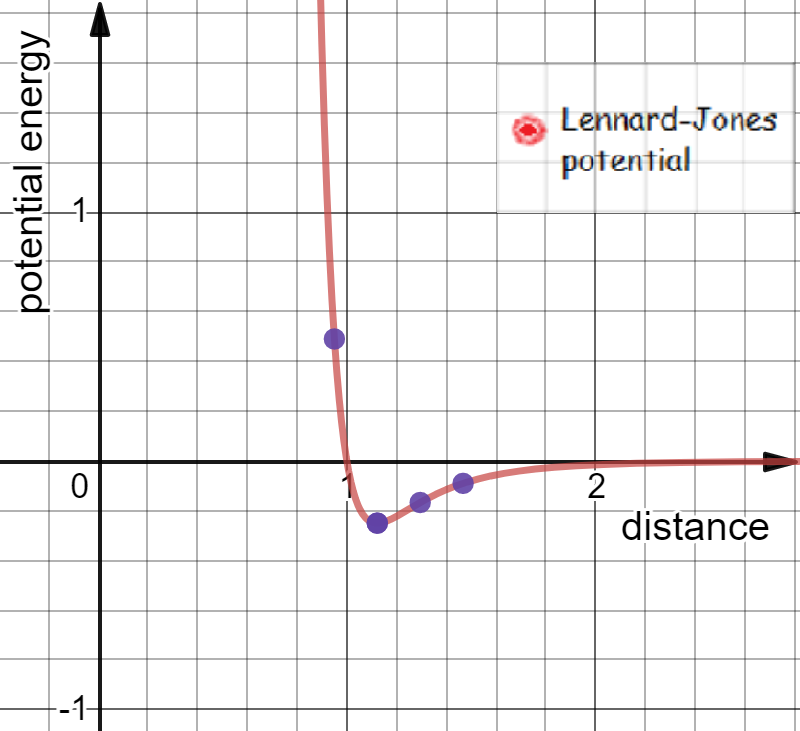}
    \caption{Lennard-Jones potential is a short range potential which is one of the best suited models for showing van der Waals interaction. As a result, it has become fundamental in modelling the molecular interaction of real substance. The points highlighted here are the mid points of region we took for discretization. The distance corresponding to the minimum potential from the origin is at x = 1.1225 and generally is called bond length for molecules.}
\label{ljp}
\end{figure}

\subsection{Discretization}
Since, the state-space is `discretized' for simulation, the Hamiltonian operator, therefore, is given as

\begin{equation}
\widehat{H}_d = {\hat{p}_{1d}}^{2} + {\hat{p}_{2d}}^{2} + ({\hat{x}_d}^{-12} -  {\hat{x}_d}^{-6})
\end{equation}

where $\hat{p}_{d}$ and $\hat{x}_{d}$ are the momentum and position operators in finite dimensions. These position and momentum operators can be expressed as in their diagonalized form which has eigenvalues as their diagonal entries in their respective space.

\subsubsection{Position operator}

\begin{figure}[H]
\centering
\includegraphics[scale=.55]{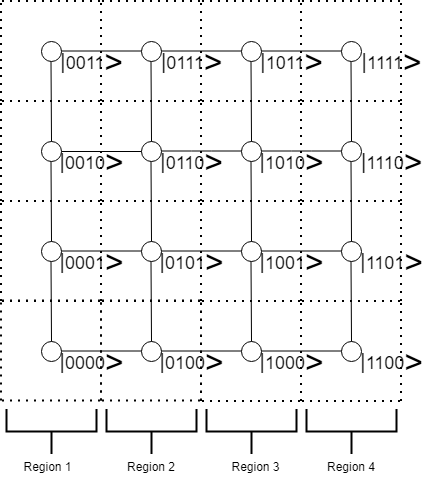}
\caption{Spatial grid: from the picture of any edge of this dotted grid as our spatial grid we can understand this state-space contains all these permutations. The potential is computed on the mid-points of each section of one-dimensional space. This is to take into account that the particle could be anywhere within a section. For instance, if we assume the origin to be at the bottom left corner, the first row tells about the states when one particle is at the origin and the other particle can occupy any of the four spatial positions and thus the state vector corresponding to that case is encoded.}
\label{spg}
\end{figure}

If we take $N$ distinct lattice points of our one dimensional discrete space, there could be $N^2$ possible position-eigenstates for this system, from the permutations. Therefore, $4$ discrete spatial points have been taken wherein they can exist. This choice of size of lattice could be any, but depending on the increasing complexity, the optimum limit is to be taken. The resolution chosen is coarse due to the technological limitations of the current quantum computers in handling a certain number of qubits. After choosing an origin we can give integral values to those positions, with respect to our length scale. As discussed before, there would be $16$ eigenstate vectors representing the spatial permutations of particles (Fig \ref{spg}).

If $X_i$ and $X_j$ are the distances from the origin of the particle, the relative distance could be written as,

\begin{equation}
    X_{ij} = |X_i-X_j| = |i-j|\Delta l
\end{equation}

where $\Delta l$ is resolution of space. Now, when $i=j$, $X_{ij} \in (0, \Delta l)$, which means the particle could be anywhere between $X_{i}+\frac{\Delta l}{2}$ and $X_{i}-\frac{\Delta l}{2}$, if other particle is fixed and vice-versa. The most suitable approximation would be to take $X_{ij} = \braket{\Delta X} = \frac{\Delta l}{2} = \frac{1}{2}$.

Therefore, the potential operator in diagonal form becomes, given the position eigenvalues

\begin{equation}
    \widehat{V} =Diag \Big\{ \big( |{i-j}|\Delta l \big)^{-12}  - \big( |{i-j}|\Delta l \big)^{-6} \Big\}
\end{equation}
and $i \ne j$ for this notation and $i,j \in \{0,1,2,3 \}$. 

\subsubsection{Kinetic Energy Operator}
The kinetic energy of a particle (having mass equals to $1/2$ for simplicity) is related to momentum as $\hat{K} = \hat{p}^2$. Since to use the diagonalized form, we first write the wave function in the momentum representation. The momentum eigenvalues of a particle in P-basis for its eigenstate $|l\rangle$ is given by \cite{tunneling},

\begin{equation}
p_l = 
     \begin{cases}
       \text{$\frac{2\pi}{2^n}l$}& \quad\text{$0\leq l \leq2^{n-1}$}\\
       \text{$\frac{2\pi}{2^n}(2^{n-1}-l)$}&\quad\text{$2^{n-1}<l<2^n$}\\
     \end{cases}
\end{equation}

where $n$ is the dimension of the space. Since for each individual particle there are $4$ possible states in position space, there will be 4 states in momentum space too. The `total' wavefunction for two independent free particles can be decomposed according to the $ \Psi = \Psi_1 \Psi_2$, where $\Psi_i$ is the wavefunction of an individual particle \cite{qsl_MalikRG2019}. Therefore, for each particle the kinetic energy operator is 

\begin{equation}\label{ko}
    \widehat{K}_l = \Big( Diag \Big\{ 0, \frac{\pi}{2},0, -\frac{\pi}{2} \Big\} \Big)^2
\end{equation}

But before operating directly to the wavefunction, we need to transform our state, using quantum Fourier transform, into momentum basis representation.

\subsubsection{Quantum Fourier Transform}
Quantum Fourier transform is a discrete Fourier transform on complex amplitudes of a quantum state. On applying to our wavefunction in X-basis, we can transform it to P-basis, so that the diagonalized momentum operator can be applied to the respective eigenfunctions in p-space. An inverse quantum Fourier transform can be done to bring back the wave function into position-space. It is a unitary operator so can be physically realizable. The general form of QFT can be given as \cite{wqft},

\begin{equation}
        \ket{\Phi} = (QFT)\ket{\Psi} = (QFT)\sum_{n=0}^{N-1} c_n \ket{n} =\sum_{k=0}^{N-1} b_k \ket{k}
\end{equation}
where 
\begin{equation*}
         b_k = \frac{1}{\sqrt{N}} \sum_{n=0}^{N-1} c_n exp \big( \frac{2 \pi \iota nk}{N} \big)  
\end{equation*}

where $N$ is the total number of bases. Therefore the momentum operation can be applied as follows
\begin{equation}
    \hat{P_x} \ket{\Psi} = (QFT)^{-1}\hat{P_l}(QFT)\ket{\Psi}
\end{equation}

where $\hat{P_x}$ is momentum operator in X-basis, $(QFT),\ (QFT)^{-1}$ refers to quantum Fourier transform operators and $\hat{P_l}$ is the momentum operator in P-basis.

The unitary matrix of QFT is given by \cite{qft},
\begin{equation}
    U_{QFT} = \frac{1}{\sqrt{N}} \sum_{x=0}^{N-1} \sum_{y=0}^{N-1} e^{2\pi \iota \frac{xy}{N}} \Ket{P}\bra{X}
\end{equation}

\subsection{Wavefunction}

After using the first order approximated Baker-Campbell-Hausdorff formula, the final unitary operation on the wave function for our system looks like as follows, and thus the time evolution can be studied.

\begin{equation} \label{last1}
    \hat{U}=e^{-\iota \hat{H}\Delta t} \approx U_{(QFT)}(e^{-\iota\hat{P}^2 \Delta t}) U_{(QFT)}^{-1}( e^{-\iota \hat{V} \Delta t})
\end{equation}

Also, any arbitrary wavefunction can be decomposed through their eigenfunctions in their respective space.

\begin{equation} \label{last}
    \Ket{\Psi(x_1,x_2,t)} = \sum_{i,j=0}^{2^n-1} \psi(x_i,x_j,t) \Ket{ij} 
\end{equation}

where $\sum_{i,j=0}^{2^n-1} |\psi(x_i,x_j,t)|^2 =1$  and  $\Ket{ij}$ is the two-dimensional kronecker delta function. Therefore, as a summary, the evolution of state is given using Eqs. \ref{last1} and \ref{last}.

\begin{equation} 
    \sum_{i,j=0}^{15} \psi(x_1,x_2,t+\Delta t) \Ket{ij}  = \Hat{U} \big[\sum_{i,j=0}^{15} \psi(x_1,x_2,t) \Ket{ij}\big]
\end{equation}

These states and operators are now mapped onto qubits and qubit-gates respectively for a simulation.

\section{Circuit Implementation}\label{qsp_sec3}

\subsubsection{Main Idea}
One of the postulates of quantum mechanics says that every quantum observable (like momentum, position, and energy) is associated with an operator, whose eigenvalues correspond to all the possible values of that observable. In a certain sense, this incorporates the physical indeterminacy with the measurement of a quantum system. Every mathematical operation and so does any physical operation can be boiled down to fundamental logic gates. Any quantum logic gate for a two-state system is realized using four fundamental gates called Pauli gates. So our dynamical system can be mapped to an electronic device by mere quantum gates \cite{qgates}. Therefore, on the basis of operations, we design the combination of gates that further acts on the set of qubits on which the wavefunction is mapped. The phase rotation is carried out by a series of CU1 gates (controlled-U1) and Hadamard gates \cite{qgates}.

Since we are simulating 4 lattice points that combine 16 dimensional Hilbert space, therefore we need 4 qubits to describe those 16 different states and only 1 more qubit is used as an ancillary qubit. Further, we optimize the circuit by reducing the redundant gates and exploiting the symmetry showed by the operators. We manage to apply the kinetic energy operator without any use of an ancillary qubit. Following this algorithm, one would need $N+1$ qubits to implement potential with most efficiency \cite{eff1, eff2} where $2^N (N\in \mathbb{N}$) is the number of lattice points. Other values of lattice points would result in wastage of qubits since n qubits can represent up to $2^N$ states. These $2^N$ qubits that together store the complete state of the system (Eq. \eqref{last}). The probability distribution of each qubit state corresponds one to one with the probability distribution of each eigenfunction after the time evolution (Eq. \eqref{eig}).

The final state in spectral decomposed form will be interpreted as 
\begin{equation} \label{eig}
     \Ket{\Psi} \rightarrow \sum_{l=0}^{2^4-1}\chi \Ket{l_b}
 \end{equation}
where $\chi$ is the amplitude of respective eigenstates of the system which is cloned by the qubit's state $\Ket{l_b}$ expressed in binary.
 
\subsubsection{Technique} \label{section}

The circuit has been coded on Qiskit platform \cite{qis}, for which the main idea is presented, through graphical circuit composer, here. The complete circuit can be segregated into 5 parts as suggested by Eq. \eqref{last1}. 

\begin{itemize}
\item initial state preparation
\item potential energy operation
\item kinetic energy operation
\item quantum Fourier transform operation
\item measurement operation 
\end{itemize}

As mentioned, each state of qubits will encode an initial state of the system. The IBM quantum experience default initial state is $\Ket{000..}$. To obtain various initial conditions, we can use NOT gates or simply Hadamard gates.
 
\begin{figure}[H]
     \centering
     \includegraphics[scale=0.8]{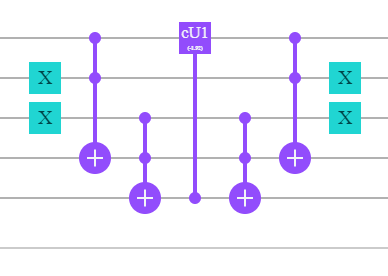}
     \caption{The filtering has been explained here with an example. The main purpose of the circuit is to operate certain operations on the $\Ket{1}$ depending upon the function it is performing. This circuit picks out the state $\Ket{001}$ with the help of Toffoli gates, then the phase-rotation has been applied to any of the qubits having state $\Ket{1}$. The mirror circuit succeeding, is to revert back the changes applied for filtering which is so that other operations could be applied. In the simulation we optimized it so as to reduce the ancillary qubits.}
     \label{filter}
\end{figure}

\begin{figure*}[]
     \centering
     \includegraphics[height=4cm, width=18cm]{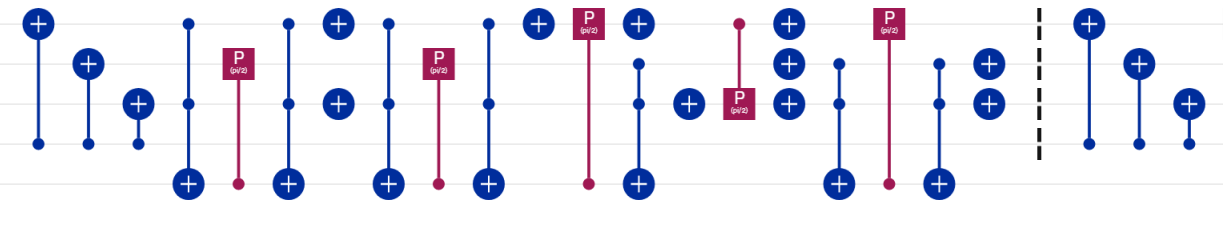}
     \caption{Graphical circuit for potential operator. The circuit is divided as given in the Section \ref{section}. The circuit is exploiting the symmetry of potential operator. The first three CNOT gates invert the three inputs which is equivalent of folding half of the table \ref{tab} coinciding it to the other half. The rest of the circuit is filtering and applying phase rotation corresponding to the state.}
     \label{potop1}
\end{figure*}

\begin{figure*}[]
\centering
\includegraphics[scale=.7]{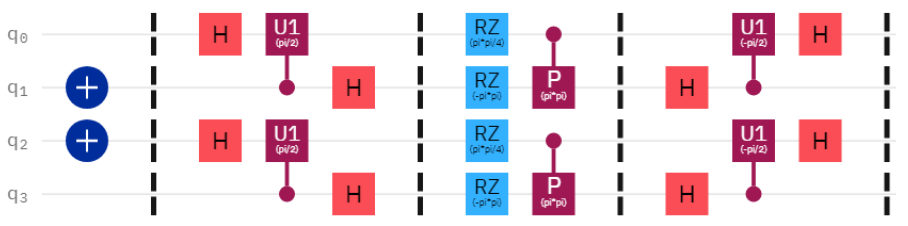}
\caption{Graphical circuit for momentum operator, Qft. The circuit is divided as given in the Section \ref{section}. As we go from right to left, we prepare an initial state, apply quantum Fourier transform, then act upon by the momentum operator, after which to revert back the changes inverse quantum Fourier transform is applied.}
\label{potop2}
\end{figure*}

For designing the circuit for the potential operator (Fig. \ref{potop1}), we took the advantage of the symmetry in the arguments (Table \ref{tab}) and the fact that phase rotation works only on $\Ket{1}$ state. For that, we used a series of Toffoli gates (two control NOT gate) to invert the remaining input and apply $cU1$ gates for a particular state using filtering technique. This technique picks a particular state and applies a phase-rotation to that state according to the unitary operator given (Fig. \ref{filter}). The final circuit has been optimized by canceling redundant gates (Fig. \ref{potop1} and Fig. \ref{potop2}). The real code includes more iterations and a smaller time-step but the figure here was given just to give an idea.

\begin{table}
     \centering
     \begin{tabular}{|c|c|c|c|}
     \hline
     States & \multicolumn{3}{|c|}{Diagonal Entries of Potential Energy operator}    \\
     \hline     
             &  &  \multicolumn{2}{|c|}{after extraction of common phase}  \\
     \hline
     & & & \\
      $\Ket{0000}$    &   0.49024 & 0        & $V_1$\\
      $\Ket{0001}$    &  -0.25    & -0.74024 & $V_2$\\
      $\Ket{0010}$    &  -0.16711175 & -0.65735175     & $V_3$\\
      $\Ket{0011}$    &  -.0901352& -0.5803752  & $V_4$\\
      $\Ket{0100}$    &  -0.25    & -0.74024 & $V_2$\\
      $\Ket{0101}$    &   0.49024 & 0        & $V_1$ \\
      $\Ket{0110}$    &  -0.25    & -0.74024 & $V_2$\\
      $\Ket{0111}$    &  -0.16711175 & -0.65735175     & $V_3$\\
      & & & \\
      \hline
      & & & \\
      $\Ket{1000}$    &  -0.16711175 & -0.65735175     & $V_3$\\
      $\Ket{1001}$    &  -0.25    & -0.74024 & $V_2$ \\
      $\Ket{1010}$    &   0.49024 & 0        & $V_1$ \\
      $\Ket{1011}$    &  -0.25    & -0.74024 & $V_2$\\
      $\Ket{1100}$    &  -.0901352& -0.5803752  & $V_4$\\
      $\Ket{1101}$    &  -0.16711175 & -0.65735175     & $V_3$\\
      $\Ket{1110}$    &  -0.25    & -0.74024 & $V_2$\\
      $\Ket{1111}$    &   0.49024 & 0        & $V_1$ \\
      & & & \\
      \hline
     \end{tabular}
     \caption{Diagonal entries of potential operator. Since the probability amplitudes are independent of the global phase, we can omit it in the expression which heavily simplifies our equations. This does not affect the relative phase of each operator. The symmetry shown, can be exploited to optimize our circuits.}
     \label{tab}
 \end{table}

The similar technique applies for designing momentum operator (Eq. \eqref{ko}). Since, it is applied to each pair of the qubits separately, it is mapped onto $R_z$ and CPhase gates which are operated individually on $\ket{q_1q_0}$ and $\ket{q_3q_2}$ (Fig. \ref{potop2}).

 \begin{figure}[H]
     \centering
     \includegraphics[scale=0.42]{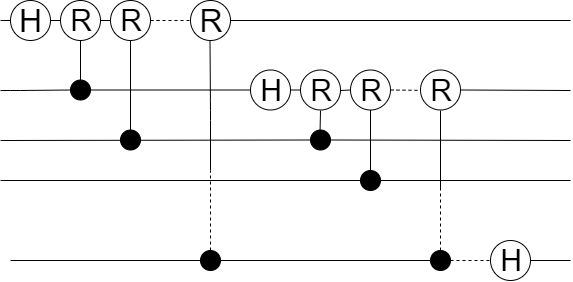}
     \caption{General n-qubit QFT circuit where H stands for Hadamard gate and R stands for the phase-rotation gate for customizing input argument.}
     \label{qft}
 \end{figure}
 
The standard circuit implementing the QFT to n-qubits is executed through a series of Hadamard and cU1 gates (controlled-phase gates) given in Fig. \ref{qft} \cite{nielsen}. The argument of each individual $c\, U1$ gate is given by,
 
 \begin{equation}
     \phi = \frac{2\pi}{2^N} \iota
 \end{equation}
 
\begin{figure}[H]
     \centering
     \includegraphics[scale=.75]{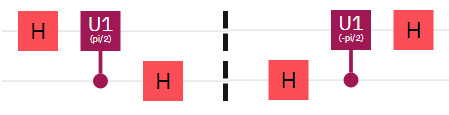}
     \caption{The circuit implementation of QFT and QFT inverse on IBM circuit composer for 2-qubit system.}
     \label{4_qft}
 \end{figure}

N being the number of qubits. The circuit for inverse QFT is mirror image of QFT with conjugate arguments (i.e., $-\phi$) (Fig. \ref{4_qft}).
 
For 2-qubit system, our circuit is represented on IBM circuit composer as given in Fig. \ref{4_qft},
 
\begin{table}[h]
     \centering
     \begin{tabular}{ |p{1cm}|p{1.5cm}|p{1.5cm}| }
     \hline
     States & \multicolumn{2}{|c|}{Diagonal Entries } \\
     \hline
        & $\Ket{K}$ & $ e^{-\iota \Ket{K} \Delta t}$ \\
    \hline    
      $\Ket{00}$    &   0   & 1  \\
      $\Ket{01}$    &  $\frac{\pi^2}{4}$ & $ e^{-\iota \frac{\pi^2}{4}\Delta t}$ \\
      $\Ket{10}$    &  0  & 1 \\
      $\Ket{11}$    &   $\frac{\pi^2}{4}$  & $ e^{-\iota \frac{\pi^2}{4}\Delta t}$ \\
      \hline
     \end{tabular}
     \caption{Phase values of Kinetic Energy operator}
     \label{tab'}
\end{table}

The last step is the measurement, where the quantum state is measured and stored in a classical register.

 \begin{figure}[H]
     \centering
     \includegraphics[scale=0.89]{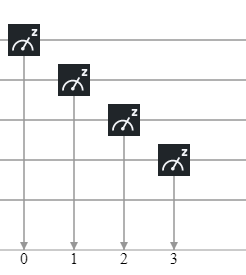}
     \caption{The final state of qubits is stored in a classical computer through registers. We need to store amplitude of each state. Since, an amplitude is a complex number, we need a space equivalent to storing 2 float numbers. Using 32 bits (4 bytes) for each floating point number, a quantum state of n = 4 qubits require 128 byte of memory which rise upto 1GByte for n=27.}
     \label{measure}
\end{figure}
 
The $\Delta t$ (time step) is taken to be 0.05 and 0.01 and convergence is checked. The final time is chosen to be $\sqrt{2}$ units calculated from the constants $\sigma, \epsilon$ and $m$.

\section{Experimental Results}\label{qsp_sec4}

\begin{figure}[h]
     \centering
     \includegraphics[scale=0.67]{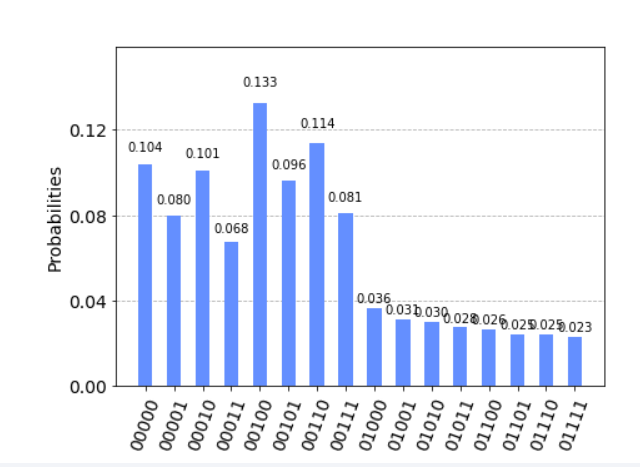}
     \caption{This is the enlarged image of the $29^{th}$ evolution of the state $\Ket{0101}$. The result is obtained through the IBM Q5 Yorktown, backened name: ibmqx2 and version: 2.2.5.  }
     \label{result4}
 \end{figure}

\begin{figure*}[t]
     \centering
     \includegraphics[scale=0.85]{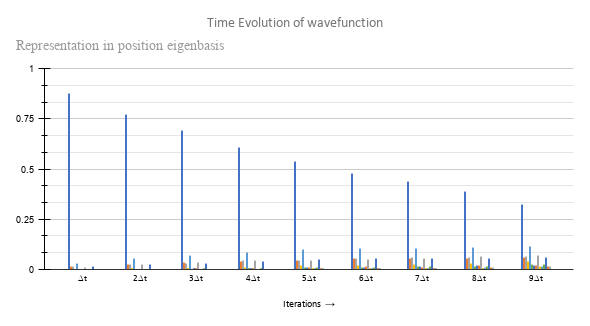}
     \caption{Time evolution of the state $\Ket{0001}$ which corresponds to classically least energy state. From the number of iterations one can infer that it is relatively stable state. The evolution is presented upto $9^{th}$ iterations in this diagram.}
     \label{result1}
\end{figure*}

\begin{figure*}[t]
\centering
\includegraphics[width=0.85\textwidth]{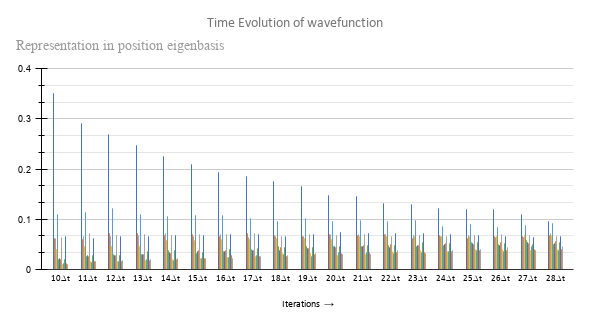}
\caption{The previous evolution is continued in this diagram. The evolution of each position eigenvector from $9^{th}$ to $28^{th}$ iteration is presented after which the states are diffused in other states comparably. The last state is again presented in Fig. \ref{result3}.}
\label{result2}
\end{figure*}

\begin{figure*}[t]
\centering
\includegraphics[scale=0.67]{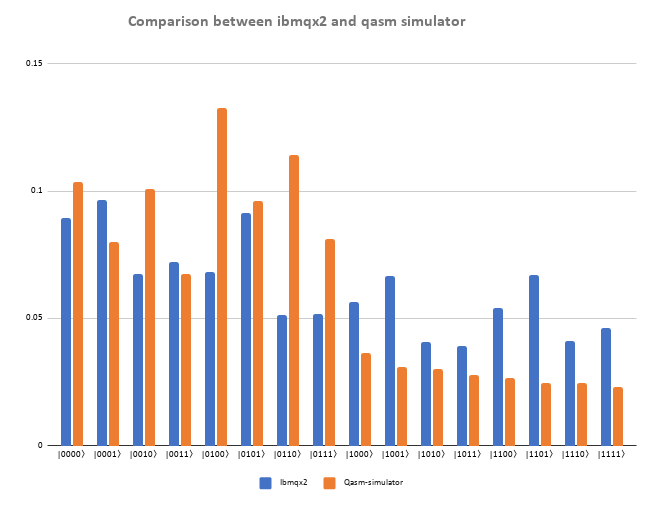}
\caption{Experimentally obtained results from ``ibmqx2" for probability distribution. Similar observations have been obtained as case ``Qasm-Simulator". The horizontal axis represents the distinct states and the vertical axis represents the corresponding probabilities. The orange bars corresponds to ``Qasm-Simulator" and the blue bars corresponds to ``Ibmqx2".}
\label{result3}
\end{figure*}

\begin{figure*}[h]
{\includegraphics[height= 2in, width = 3in]{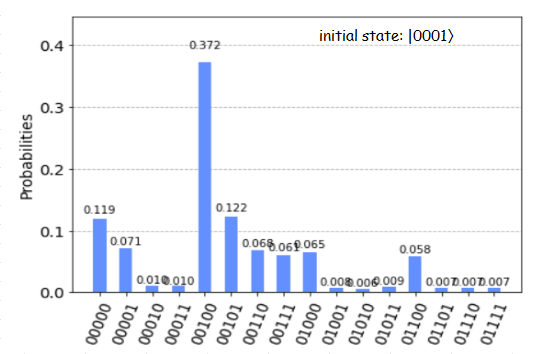}}\hfill 
{\includegraphics[height= 2in, width = 3in]{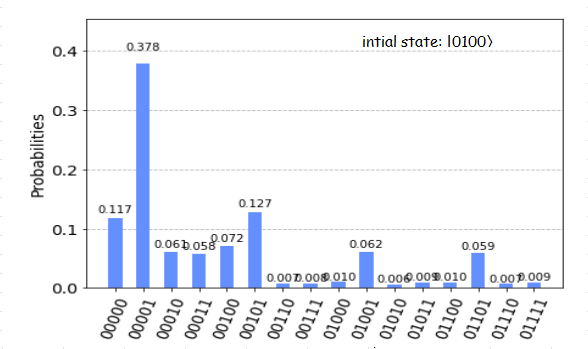}}\hfill
\\[\smallskipamount]
{\includegraphics[height= 2in, width = 3in]{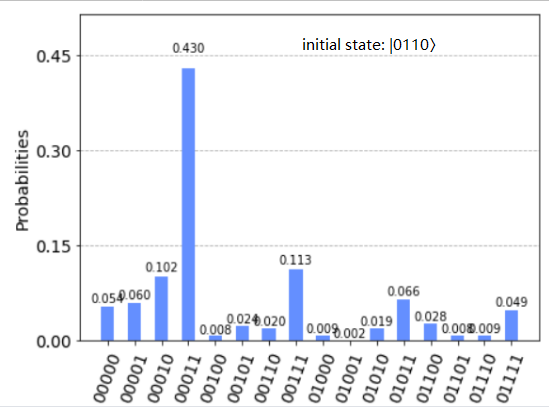}}\hfill
{\includegraphics[height= 2in, width = 3in]{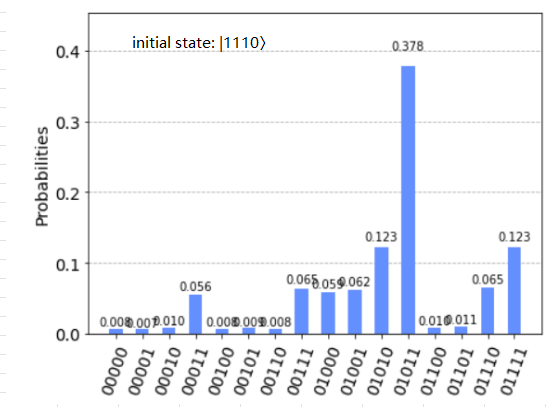}} \hfill
\caption{Final state after 29$^{th}$ iterations for different initial states showing similar behaviour is given. The results were obtained by Qasm-Simulator. The various states are $\Ket{0001},\ \Ket{0100},\  \Ket{0110}$ and  $\Ket{1110}$ respectively.}
\label{some example}
\end{figure*}

\begin{figure*}[h]
{\includegraphics[height= 2in, width = 3in]{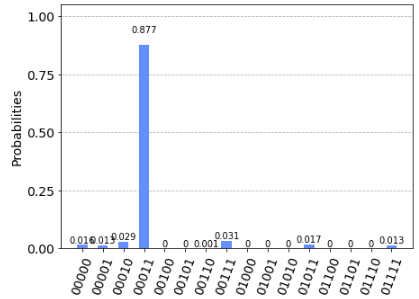}}\hfill 
{\includegraphics[height= 2in, width = 3in]{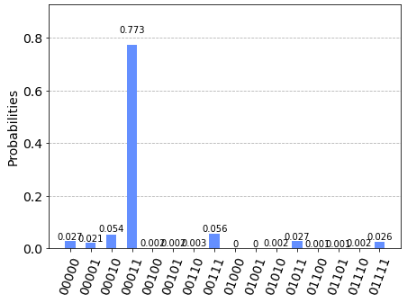}}\hfill
\\[\smallskipamount]
{\includegraphics[height= 2in, width = 3in]{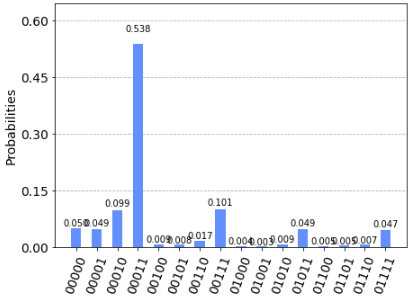}}\hfill
{\includegraphics[height= 2in, width = 3in]{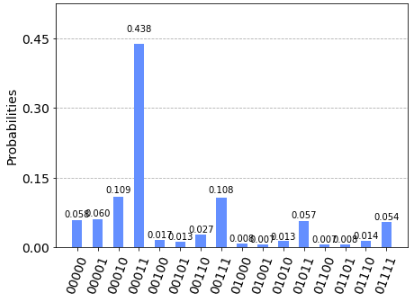}} \hfill
\\[\smallskipamount]
{\includegraphics[height= 2in, width = 3in]{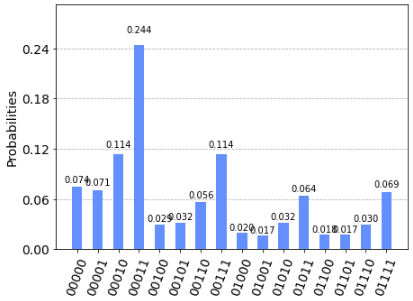}} \hfill
{\includegraphics[height= 2in, width = 3in]{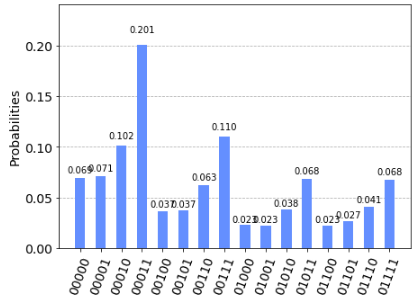}} \hfill
\\[\smallskipamount]
{\includegraphics[height= 2in, width = 3in]{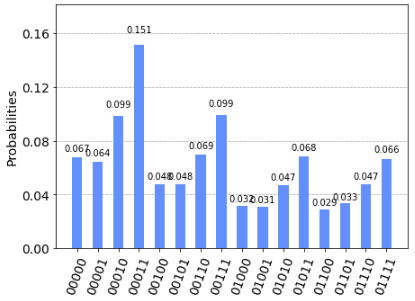}} \hfill
\caption{The following are snapshots of discrete time steps of evolution of state $\Ket{0011}$. The snapshots are for $1^{st},\ 3^{rd},\ 5^{th},\ 7^{th},\ 13^{th},\ 16^{th}$ and $20^{th}$ time-step. You can see the pattern emerging which gives the higher probability to other states with same classical energy. Apart from this you can see that it diffused to other states at a higher rate than other initial condition.}
\label{some other example}
\end{figure*}

In order to perform the simulation, various time-steps have been taken with different initial conditions to study the system. The states $\Ket{0001},\ \Ket{0100},\ \Ket{0110},\ \Ket{1001},\ \Ket{1011}$, and $\Ket{1110}$ are behaviourally similar points and correspond to classical minimum energy. The evolution of $ \Ket{0001}$ has presented here in Fig. \ref{result1} and Fig. \ref{result2}, upto a natural time scale of the system. The state being classically most favourable, diffuses to other states having peaks near the states corresponding to the same classical least energy. To compare the simulated result from ``IBM Q QASM simulator", a similar simulation is performed on quantum processor ``ibmqx2" \cite{ibmqx} which is depicted in Fig. \ref{result3}. In both the cases, the number of shots taken were 8192. In Fig. \ref{result4}, the final state after $29^{th}$ iteration is obtained from Qiskit platform, and though this single snapshot is insufficient to capture the whole process, it can be seen that the wavefunction is slowly diffusing out to other states showing increasing position uncertainty with time. To see the resemblance, we simulated the states $ \Ket{0001},\ \Ket{0100},\  \Ket{0110}$ and  $\Ket{1110}$ and have shown the wavefunction after 29$^{th}$ iteration in Fig. \ref{some example}. Similar behavior is obtained, which verifies the symmetry of the space. In contrast to this, we can see for the state $\Ket{0011}$, the wavefunction diffused to other states rather quickly in Fig. \ref{some other example}. Snapshots in an irregular range are given, which finally gives the wave function a form with a maximum probability in the same classical energy-like states.  

Ultimately, this simulation can be generalized for arbitrary spatial dimensions and the number of interacting particles. The simulation can also be carried out using an arbitrarily large number of qubits to make it more precise. The same type of optimization can be done by exploiting the symmetries of eigenvectors.

\section{Discussion and Conclusion}\label{qsp_sec5}
We successfully simulated a two-particle system in a single-dimensional grid using short-range interaction. We achieved a recognizable behavior of the interaction within the framework of 5 qubits. The results shown in Fig. \ref{result3} indicates that simulations are not appropriate for closed systems because of the characteristically different results. Since qubits are in noisy environment, we would simulate Hamiltonian which are not coherence preserving \cite{coherence}. Although the results achieved are radically (Fig. \ref{some example}, Fig. \ref{result4}) different than one would expect them to be, classically, which implies the need for a more refined grid. As the characteristic behavior is only noticeable in the short range of particles, we can increase the grid size without increasing the interactions. Similarly, we would like to simulate this in 3-dimensional space using 3-D meshing techniques. The accuracy of results will depend on the time step whose suitable value needs to be computed \cite{lloyd}. These amendments can only be made once we develop ways to process more qubits. After that, we hypothesize that we can simulate multibody quantum systems and delve deeper into the statistical nature of quantum systems. 

\clearpage

\pagebreak

\section*{Acknowledgements}\label{qsp_sec6}
Prabhat would like to thank Bikash's Quantum (OPC) Private Limited for providing hospitality during the course of the project work. Prabhat would like to thank all the colleagues and his mentor B.K.B. who supported him through the project. The authors acknowledge the support of IBM Quantum Experience. The views expressed are those of the authors and do not reflect the official policy or position of IBM or the IBM Experience team.

\section*{Author Contributions}
Theoretical analysis, simulation of a quantum circuit, collection, and analysis of data were done by Prabhat. The design of quantum circuits was done by Prabhat and B.K.B. The project was supervised by B.K.B. B.K.B. has thoroughly reviewed the manuscript.

\iffalse
\newpage
\section*{Appendix}\label{qsp_sec7}
We are attaching the code here. 
\lstinputlisting[language=Python]{Final Code.py} 
\fi

\end{document}